# What makes papers visible on social media? An analysis of various document characteristics[1]

Zohreh Zahedi[*], Rodrigo Costas[*], Vincent Larivière[**] and Stefanie Haustein[**]

[*] z.zahedi.2@cwts.leidenuniv.nl; rcostas@cwts.leidenuniv.nl
Centre for Science and Technology Studies (CWTS), Leiden University, Wassenaarseweg 62A, Leiden, 2333 AL (The Netherlands)

[**] vincent.lariviere@umontreal.ca; stefanie.haustein@umontreal.ca
École de bibliothéconomie et des sciences de l'information, Université de Montréal, C.P. 6128, Succ. Centre-Ville, Montréal, H3C 3J7 (Canada)

**ABSTRACT**
In this study we have investigated the relationship between different document characteristics and the number of Mendeley readership counts, tweets, Facebook posts, mentions in blogs and mainstream media for 1.3 million papers published in journals covered by the Web of Science (WoS). It aims to demonstrate that how factors affecting various social media-based indicators differ from those influencing citations and which document types are more popular across different platforms. Our results highlight the heterogeneous nature of altmetrics, which encompasses different types of uses and user groups engaging with research on social media.

**INTRODUCTION**
Five years after the introduction of the term (Priem, et al. 2010), altmetric indicators can be found on most of major publishers platforms, and are increasingly used in research evaluation (Wilsdon et al., 2015). Although some factors such as document age (Thelwall, et al, 2013), discipline (Haustein, Costas, & Larivière, 2015), topic (Costas, Zahedi, & Wouters, 2015) as well as countries (Alperin, 2015), have been shown to affect the various indicators, the processes which make a scientific paper visible on social and mainstream media are still not yet fully understood. Haustein et al. (2015) showed that factors which typically influence citations counts had a smaller or opposite effect on social and mainstream media mentions and that the usage pattern differed in particularly regarding document types. This study builds upon this work, taking into account a longer citation and social media window and expanding it by Mendeley readership counts. It addresses the following research questions:

What is the effect of document characteristics on the number of Twitter, Facebook, blogs and mainstream media mentions as well as on Mendeley readership counts? Particularly,
  1. How do these effects compare with that observed for citations?
  2. How do these effects differ across document types?

---

[1] This work was supported by the Alfred P. Sloan Foundation Grant #2014-3-25, Leiden University Fund (LUF) Grant # 4509/22-1-14\T, Vw and by funding from the DST-NRF Centre of Excellence in Scientometrics and STI Policy (South Africa).

## DATA AND METHODS

This study builds upon and expands the analysis by Haustein et al. (2015) and compares the number of tweets, public Facebook posts, mentions in blogs and mainstream media, and Mendeley readership counts to citations received by WoS publications with a DOI published in 2012 (n=1,339,279). Citations from the CWTS in-house database were considered until September 2015 and altmetrics were collected in July 2015, expanding the windows used by Haustein et al. (2015). Twitter [T], Facebook [F], blogs [B] and mainstream media [M] mentions were obtained from Altmetric.com and Mendeley readership counts [MR] were collected using the Mendeley REST API.

The analyzed document properties included the document type as indicated by WoS [DT], the number of pages [PG], cited sources in the reference list (including non-source items) [NR], and characters in the title [TI], as well as number of authors [AU], institutions [IN] and countries [CU] of the paper. The percentage of papers with at least one citation or social and mainstream media event count (coverage), the average number of counts per paper (density) and the average number of counts for documents with at least one count (intensity) were computed. Correlations are based on Spearman's $\rho$.

## RESULTS AND DISCUSSION

Table 1 shows that slightly more papers had been saved to Mendeley (84.2%) than cited (81.7%). For other social media platforms, coverage is much lower, with 22.6% of papers receiving at least one tweet, 5.2% being shared publicly on Facebook, 2.3% mentioned in blog posts, and 1.1% discussed by mainstream media. Reviews and articles are the document types that were most commonly cited or saved on Mendeley, while editorial material and news items were particularly popular on Twitter, Facebook, blogs, and mainstream media. Although both coverage and density were higher for reviews and articles, editorials and news items were also frequently saved by Mendeley users.

**Table 1.** Prevalence (coverage in %, density, intensity) of citations and social media metrics per document type.

|  |  | All document types | Article | Biographical Item | Book Review | Correction | Editorial Material | Letter | Meeting Abstract | News Item | Review |
|---|---|---|---|---|---|---|---|---|---|---|---|
| N |  | 1,338,885 | 1,132,428 | 2,302 | 21,710 | 9,817 | 60,533 | 29,410 | 13,071 | 4,880 | 64,734 |
| % |  | 99.97 | 84.56 | 0.17 | 1.62 | 0.73 | 4.52 | 2.2 | 0.98 | 0.36 | 4.83 |
| Citations | Coverage | 81.72% | 86.89% | 9.86% | 2.13% | 49.53% | 49.53% | 47.49% | 6.34% | 37.34% | 94.14% |
| Citations | Density | 7.68 | 7.84 | 0.17 | 0.03 | 0.44 | 2.26 | 1.69 | 0.09 | 1.39 | 18.78 |
| Citations | Intensity | 9.40 | 9.02 | 1.70 | 1.23 | 2.31 | 4.57 | 3.56 | 1.35 | 3.72 | 19.95 |
| Blogs | Coverage | 2.28% | 2.27% | 0.70% | 0.18% | 2.21% | 2.71% | 1.00% | 0.05% | 2.68% | 3.85% |
| Blogs | Density | 0.04 | 0.04 | 0.01 | 0.00 | 0.03 | 0.05 | 0.02 | 0.00 | 0.03 | 0.06 |
| Blogs | Intensity | 1.78 | 1.80 | 1.06 | 1.10 | 1.21 | 1.77 | 1.78 | 1.00 | 1.27 | 1.62 |
| Twitter | Coverage | 22.55% | 21.98% | 13.99% | 5.42% | 10.68% | 28.57% | 19.02% | 2.21% | 47.97% | 38.67% |
| Twitter | Density | 1.02 | 0.94 | 0.41 | 0.11 | 0.20 | 2.05 | 0.59 | 0.05 | 4.26 | 1.95 |
| Twitter | Intensity | 4.52 | 4.30 | 2.90 | 1.98 | 1.90 | 7.17 | 3.12 | 2.46 | 8.89 | 5.05 |
| Facebook | Coverage | 5.20% | 4.94% | 3.00% | 1.31% | 1.33% | 7.99% | 3.64% | 0.27% | 11.13% | 7.99% |

|  |  |  |  |  |  |  |  |  |  |  |
|---|---|---|---|---|---|---|---|---|---|---|
|  | Density | 0.12 | 0.11 | 0.00 | 0.02 | 0.02 | 0.18 | 0.07 | 0.00 | 0.21 | 0.25 |
|  | Intensity | 2.22 | 2.21 | 1.45 | 1.21 | 1.20 | 2.28 | 1.98 | 1.11 | 1.90 | 2.46 |
| Mainstream media | Coverage | 1.08% | 1.10% | 0.13% | 0.00% | 0.12% | 1.15% | 0.31% | 0.01% | 0.68% | 1.67% |
|  | Density | 0.02 | 0.02 | 0.00 | 0.00 | 0.00 | 0.02 | 0.01 | 0.00 | 0.01 | 0.03 |
|  | Intensity | 2.02 | 2.05 | 1.00 | 2.00 | 1.50 | 1.97 | 3.14 | 1.11 | 1.91 | 1.66 |
| Mendeley readership | Coverage | 84.23% | 87.70% | 28.37% | 28.83% | 25.37% | 68.23% | 58.36% | 24.74% | 59.41% | 93.57% |
|  | Density | 11.00 | 10.94 | 0.79 | 1.51 | 3.41 | 6.47 | 2.76 | 0.57 | 4.98 | 27.29 |
|  | Intensity | 13.06 | 12.47 | 2.80 | 5.23 | 13.46 | 9.48 | 4.73 | 2.31 | 8.38 | 29.16 |

Correlations show that Mendeley readership has the highest positive correlation ($\rho = .585$, Table 2) with citation counts, followed by Twitter ($\rho = .279$) and blogs ($\rho = .159$), while Facebook ($\rho = .142$) and mainstream media ($\rho = .115$) show positive but low correlations with citations. These findings point to different audiences and engagements on these social media platforms. While the stronger relationship between citations and readership counts likely reflect Mendeley's use in a pre-citation context (Mohammadi, Thelwall, & Kousha, 2015), the lower correlations with Twitter might be related to Twitter's inclusion of non-academic audiences. Facebook is mostly used for private rather than professional purposes (Van Noorden, 2014), and users generally interact in closed rather than open groups. Blogs and mainstream media are very selective in the sense that only a fraction of papers are mentioned. It should be noted that the low correlations are largely caused by low coverage: more than 98% of papers did not get mentioned in blogs or mainstream media. Both of these sources are targeted at larger audiences than scientific papers and are generally written in a less technical language, while blogs mainly focus on academia and mainstream media target a general audience. It should also be mentioned that papers covered by mainstream media and blogs are often published in multidisciplinary scientific journals such as *Nature* or *Science* (Costas et al, 2015).

**Table 2.** Correlation between document characteristics, citations and social media mentions.

|  | PG | NR | TI | AU | IN | CU | C | MR | B | T | F | M |
|---|---|---|---|---|---|---|---|---|---|---|---|---|
| **PG** | 1.000 | 0.622 | 0.079 | -0.006 | 0.116 | 0.131 | 0.250 | 0.287 | 0.007 | 0.036 | 0.013 | -0.001 |
| **NR** |  | 1.000 | 0.165 | 0.155 | 0.168 | 0.146 | 0.485 | 0.471 | 0.061 | 0.145 | 0.068 | 0.043 |
| **TI** |  |  | 1.000 | 0.323 | 0.135 | 0.038 | 0.169 | 0.080 | -0.033 | -0.007 | -0.012 | -0.022 |
| **AU** |  |  |  | 1.000 | 0.494 | 0.252 | 0.320 | 0.168 | 0.031 | 0.085 | 0.047 | 0.033 |
| **IN** |  |  |  |  | 1.000 | 0.560 | 0.215 | 0.177 | 0.049 | 0.102 | 0.061 | 0.042 |
| **CU** |  |  |  |  |  | 1.000 | 0.170 | 0.153 | 0.045 | 0.060 | 0.039 | 0.036 |
| **C** |  |  |  |  |  |  | 1.000 | 0.585 | 0.140 | 0.220 | 0.120 | 0.108 |
| **MR** |  |  |  |  |  |  |  | 1.000 | 0.159 | 0.279 | 0.142 | 0.115 |
| **B** |  |  |  |  |  |  |  |  | 1.000 | 0.211 | 0.193 | 0.297 |
| **TW** |  |  |  |  |  |  |  |  |  | 1.000 | 0.328 | 0.161 |
| **FB** |  |  |  |  |  |  |  |  |  |  | 1.000 | 0.182 |
| **M** |  |  |  |  |  |  |  |  |  |  |  | 1.000 |

(N=1,339,279); PG=Page, NR= Number of References, TI= Title length, AU=Author, IN=Institute, CU=Country; C=Citations, MR=Mendeley readership counts, B=Blogs, T=Twitter, F=Facebook, M=Mainstream Media

At the level of document characteristics, Mendeley readership counts exhibit the highest positive correlation with the number of references made ($\rho = .471$) — showing trends similar to citations — followed by number of pages ($\rho=.287$) and title length ($\rho=.080$). However, the latter does not seem to have a large effect on attracting Mendeley users. Readership count patterns are comparable to citations. Although correlations were low, negative correlations between the other metrics and title length (as well as document length for main stream media) suggests that social media users, to the opposite of citing authors, exhibit a preference for short titles (and documents length). The highest correlation of citations is with the number of authors ($\rho=0.320$), followed by number of institutions ($\rho=0.215$) and countries ($\rho=0.170$). Altmetrics show less pronounced effects regarding these collaboration indicators slightly different and less pronounced effects.

**PRELIMINARY CONCLUSIONS AND OUTLOOK**

This paper provided insights on the relationship between social and mainstream media visibility and various documents characteristics. It is shown that some of them influence the extent to which they are cited or shared on social media. However, patterns vary between indicators. While Twitter, Facebook, blogs and mainstream media mentions are different from citations as reflected in low correlations and the popularity of so-called "non-citable" document types, Mendeley exhibits patterns similar to citations, which is likely due to its use in a pre-citation context. Our results thus highlight the heterogeneous nature of altmetrics, which encompasses different types of uses and user groups engaging with research on social media. Future research will include to what extent this pattern is different across disciplines as well as how these indicators change by different levels of collaboration and document characteristics by applying multiple regression analysis.